\documentclass[a4paper, 11pt]{article}
\date{December 2005}
\usepackage{epsfig}
\usepackage{latexsym}
\usepackage{amssymb}
\newcommand{\be}{\begin{equation}}
\newcommand{\ee}{\end{equation}}
\newcommand{\ba}{\begin{eqnarray}}
\newcommand{\ea}{\end{eqnarray}}
\newcommand{\bi}{\begin{itemize}}
\newcommand{\ei}{\end{itemize}}
\newcommand{\tr}{{\rm Tr\,}}
\newcommand{\re}{\mathop{\rm Re}}

\newcommand{\nn}{\nonumber \\}

\newcommand{\quarter}{{\textstyle\frac{1}{4}}}

\newcommand{\threequarter}{{\textstyle\frac{3}{4}}}

\newcommand{\ovl}{\overline}

\newcommand{\<}{\langle}
\renewcommand{\>}{\rangle}
\newcommand{\eq}{Eq.~}

\newcommand{\la}{\label}

\newcommand{\betas}{\beta_{\sigma}}
\newcommand{\betat}{\beta_{\tau}}
\newcommand{\as}{a_{\sigma}}
\newcommand{\at}{a_{\tau}}
\newcommand{\fp}{f_{+}}
\newcommand{\Fp}{F_{+}}

\newcommand{\Zps}{Z^{+}_{\sigma}}
\newcommand{\Zpt}{Z^{+}_{\tau}}
\newcommand{\Zms}{Z^{-}_{\sigma}}
\newcommand{\Zmt}{Z^{-}_{\tau}}
\newcommand{\Zp}{Z^{+}}
\newcommand{\Zm}{Z^{-}}

\newcommand{\fm}{f_{-}}
\newcommand{\Fm}{F_{-}}
\newcommand{\fpm}{f_{\pm}}
\newcommand{\Fpm}{F_{\pm}}
\newcommand{\txts}{\textstyle}

\newcommand{\Sm}{S_{-}}
\newcommand{\Sp}{S_{+}}
\newcommand{\Ss}{S_{\sigma}}
\newcommand{\St}{S_{\tau}}
\newcommand{\Nt}{N_{\tau}}
\newcommand{\Ns}{N_{\sigma}}

\newcommand{\dBdloga}{\frac{d\beta}{d\log a}}
\newcommand{\Ntref}{N_\tau^{\rm ref}}
\newcommand{\Tref}{T_{\rm ref}}
\newcommand{\dBsdlogas}{\frac{\partial\beta_{\sigma}}{\partial\log a_{\sigma}}}
\newcommand{\dBtdlogas}{\frac{\partial\beta_{\tau}}{\partial\log a_{\sigma}}}
\newcommand{\dBsdlogxi}{\frac{\partial\beta_{\sigma}}{\partial\log \xi}}
\newcommand{\dBtdlogxi}{\frac{\partial\beta_{\tau}}{\partial\log \xi}}

\newcommand{\dFdBs}{\frac{\partial F}{\beta_\sigma}}
\newcommand{\dFdBt}{\frac{\partial F}{\beta_\tau}}

\begin{document}
\begin{titlepage}
MIT-CTP 3903 \\

\begin{centering}
\vfill

 \vspace*{2.0cm}
{\bf \Large Finite Temperature Sum Rules in Lattice Gauge Theory}
 
\vspace{3.0cm}
{\bf Harvey~B.~Meyer}
\centerline{Center for Theoretical Physics}
\centerline{Massachusetts Institute of Technology}
\centerline{Cambridge, MA 02139, U.S.A.}
\vspace{0.1cm}\\
\centerline{meyerh@mit.edu}

\vspace*{4.5cm}

\end{centering}
\centerline{\bf Abstract}
\vspace{0.1cm}
\noindent 
We derive non-perturbative sum rules in SU($N$) lattice gauge theory at 
finite temperature. They relate the susceptibilities of the trace anomaly and 
energy-momentum tensor to temperature derivatives of the thermodynamic potentials.
Two of them have been derived previously in the continuum
and one is new. In all cases, at finite latttice spacing there are important corrections
to the continuum sum rules that are only suppressed by the bare coupling $g_0^2$.
We also show how the discretization errors affecting the
thermodynamic potentials can be controlled by computing these susceptibilities.
\vspace{2.0cm}
\vfill
\end{titlepage}

\setcounter{footnote}{0}
\section{Introduction\la{sec:intro}}
Sum rules in continuum QCD at zero temperature were introduced 
by Novikov et al.~\cite{novikov} and a lot of hadron phenomenology
was subsequently based on them. A few years later Michael derived sum rules for SU($N$) 
pure gauge theories in lattice regularization~\cite{michael2,rothe1,michael1},
which were only recently generalized to Wilson lattice QCD in~\cite{hmsr1}.
Since sum rules are relations that hold non-perturbatively, 
the lattice regularization provides a framework  in which
their derivation proceeds in a particularly rigorous way:
it only involves operations on multi-dimensional integrals. 

On the lattice the simplest 
identities relate zero-momentum three-point functions to 
the spectrum of the theory. By comparing the sum rules to 
continuum relations~\cite{ji}, one realizes~\cite{rothe2,gluex} that they
relate the normalization of a particular discretization of 
the trace anomaly and the energy-momentum tensor to 
anisotropy coefficients.
The latter are derivatives
of the bare lattice parameters with respect 
to physical parameters such as 
the lattice spacing in hadronic units or the ratio of spatial and 
temporal lattice spacings. Indeed this normalization is non-trivial 
since translation invariance is broken down to a discrete group 
at finite lattice spacing. 

Ellis et al. derived finite-temperature sum rules
in pure gauge theories~\cite{kapu1} and in full QCD~\cite{kapu2}.
In this paper we rederive the SU($N$) gauge theory sum rules, focusing on those 
concerning two-point functions of the trace anomaly and the 
energy-momentum tensor, in lattice regularization. We find that 
they have important corrections to the continuum versions, 
which are only suppressed by one power of the bare coupling $g_0^2$.
From the point of view of Monte-Carlo simulations, where 
thermodynamics calculations are performed around $g_0\approx 1$,
they can thus not be neglected.
We also derive a new sum rule involving only the traceless part of the 
energy-momentum tensor,
and we relate the results obtained to contact terms in the two-point
functions of the Hamiltonian.

The $p=0$ two-point function of the trace anomaly, in other words
its susceptibility, is related to the rate of change 
of $(\epsilon-3P)/T^4$ with temperature~\cite{kapu1} ($\epsilon$
is the energy density and $P$ the pressure). Because
the bulk viscosity is related to this two-point function by 
a Kubo formula~\cite{jeon}, 
it was argued recently~\cite{dima} that the bulk viscosity 
rises sharply just above the deconfining temperature $T_c$.
Direct calculations of the two-point functions at ${\bf p}=0$ 
and general $\omega=p_0$ have confirmed the existence of this effect~\cite{hmbulk}.
In the context of such calculations, the sum rule can be used to 
constrain the reconstruction of the spectral function $\rho(\omega)$.

Another application of these considerations to finite-temperature
Monte Carlo simulations is to compute directly the leading lattice spacing
dependence of the thermodynamic potentials. This idea has the most potential
of being useful in the context of full QCD simulations, where the 
computational cost is high and 
grows with a large power of the inverse lattice spacing.

Decomposing the energy-momentum tensor $T_{\mu\nu} $
into a traceless part $\theta_{\mu\nu}$ and a scalar part $\theta$ via
$T_{\mu\nu} = \theta_{\mu\nu} + \quarter \delta_{\mu\nu} \theta $,
the explicit Euclidean expressions are
\be
\theta(x) \equiv  \beta(g)/(2g) ~ F_{\rho\sigma}^a(x)  F_{\rho\sigma}^a(x)  
\qquad\qquad
\theta_{\mu\nu}(x) \equiv 
{\txts\frac{1}{4}}\delta_{\mu\nu}F_{\rho\sigma}^a F_{\rho\sigma}^a
   - F_{\mu\alpha}^a F_{\nu\alpha}^a .
\ee
The beta-function is defined by $qd\bar g/dq=\beta(\bar g)=-\bar g^3(b_0+b_1\bar g^2+\dots)$
and $b_0=11N/(3(4\pi)^2)$, $b_1=34N^2/(3(4\pi)^4 )$ in the SU($N$) pure gauge theory.
The gauge action reads $\quarter F_{\mu\nu}^a F_{\mu\nu}^a$ in this notation.
If $\<\dots\>_T$ denotes the thermal average at temperature $T$,
\be
\epsilon-3P = \<\,\theta\,\>_T -\<\,\theta\,\>_0, 
  \qquad\qquad
\epsilon+P = {\txts\frac{4}{3}}  \<\,\theta_{00}\,\>_T  ~.\la{eq:basic}  
\ee

In section 2 we introduce our notation, review the relations relevant 
to thermodynamics and introduce new anisotropy coefficients.
In section 3 we derive the sum rules on the lattice.
In section 4 we take the extreme continuum limit, $g_0^2\ll1$
and compare our results to those of~\cite{kapu1}.
Section 5 describes the possibility of computing the leading 
discretization errors affecting $\epsilon$ and $P$ in
numerical simulations, and section 6 contains concluding remarks.

\section{Thermodynamics and the energy-momentum tensor}
We consider a Euclidean lattice of  spatial extent $\Ns$ sufficiently large that 
the thermodynamics limit has been reached. The time-extent $\Nt=L_0/a$ 
fixes the temperature $T=1/L_0$. Thermal averages are denoted by $\<\dots\>$.
The temperature dependence is made explicit
by $\<\dots\>_T$ to distinguish this average from the average $\<\dots\>_0$ on a 
zero-temperature lattice. The contents of this section is mostly a review of the 
early lattice thermodynamics articles~\cite{Engels:1981qx,Karsch:1982ve}, with the  exception
of subsection~\ref{sec:dZ}. We however emphasize a lot more the role of 
$\theta_{00}$ and $\theta$, since they are the operators of interest in 
the sum rules.

\subsection{Isotropic lattice}
We start from the Wilson action~\cite{wilson74} for SU($N$) gauge theories:
\ba
S_{\rm g} &=& \beta \sum_x\sum_{\mu<\nu} S_{\mu\nu}(x), \la{eq:Sg} \\
S_{\mu\nu}(x) &=& \textstyle{\frac{1}{N}}
\re\tr\{1-U_{\mu}(x)U_\nu(x+a\hat\mu)U_\mu(x+a\hat\nu)^{-1}U_\nu(x)^{-1}  \}
\ea
and $\beta\equiv \frac{2N}{g_0^2}$.
It is useful to consider two separate sets of parameters:
\ba
\textrm{bare parameters:}&\qquad& \beta,~\Nt \\
\textrm{physical parameters:}&\qquad& a(\beta),~T(\beta,\Nt)
\ea
where  $1/T=L_0 = \Nt a(\beta)$.
With the notations
\be
S_\pm =      \Ss \pm \St, \qquad \Ss= \sum_{k<l} S_{kl}, \qquad \St =   \sum_{k} S_{0k} ,
\ee
we use the following discretizations:
\be
\Theta(x) 
= \Zp(\beta) ~\Sp,\qquad \qquad
\Theta_{00}(x) 
= \Zm(\beta) ~ \Sm,
\ee
with 
\be
\Zp(\beta) =  \frac{d\beta}{d\log a }
\qquad \qquad
\Zm(\beta) = \beta Z(\beta) .
\ee
The presence of the normalization factor $Z(\beta)=1+{\rm O}(1/\beta)$ 
must be expected, since $\int d^3{\bf x}\,\theta_{00}(x)$ is not a Noether charge, 
due to the lack of continous translation invariance on the lattice.
A precise expression can be given for $Z(\beta)$ in terms of derivatives with respect
to the anisotropy, see \eq\ref{eq:Zb} and also \eq\ref{eq:Zb2}. For a parametrization of $Z(\beta)$
in the case of SU(3) based on the data of~\cite{karsch-aniso}, see \cite{hmshear} (Eq. 6).
The continuum limit takes the form
\ba
\Fp(\beta,\Nt)  \equiv  \Nt^4(\< \Theta\>_T -\<\Theta\>_{0} )   
&\stackrel{a\rightarrow0}{\rightarrow}  &  \frac{\epsilon-3P}{T^4} \equiv \fp(T)
\la{eq:Fp}  \\
\Fm(\beta,\Nt) \equiv  \frac{4}{3} \Nt^4 \< \Theta_{00} \> 
&\stackrel{a\rightarrow0}{\rightarrow} & 
 \frac{\epsilon+P}{T^4} \equiv \fm(T),
\la{eq:Fm}
\ea
where the leading corrections are ${\rm O}(a^2)$.

\subsection{Anisotropic lattice}
On the anisotropic lattice with spatial lattice spacing $\as$ and temporal lattice spacing $\at$,
the action reads
\be
S_{\rm g} =   \sum_x \betas \Ss(x)  + \betat \St(x).
\ee
The two separate sets of parameters are:
\ba
\textrm{bare parameters:}&\qquad& \betas,~ \betat,~\Nt \\
\textrm{physical parameters:}&\qquad& \as(\betas,\betat),~\xi(\betas,\betat),~T(\betas,\betat,\Nt)
\ea
where
\ba
\xi &\equiv& \as/\at  \\
1/T=L_0 &= &\Nt \at = \Nt \as\xi^{-1}.
\ea
Obviously, $\xi=1$ when $\betas=\betat$.

We use the following discretizations:
\ba
\xi^{-3}~\Theta(x) &=& \Zps(\betas,\betat) \Ss + \Zpt(\betas,\betat) \St  , \\
\xi^{-3}~\Theta_{00}(x) &=& \Zms(\betas,\betat) \Ss - \Zmt(\betas,\betat)\St   ,
\ea
where at the symmetric point $\xi=1$,
\be
\Zps(\beta,\beta)=\Zpt(\beta,\beta) = \Zp(\beta), \qquad
\Zms(\beta,\beta) =  \Zmt(\beta,\beta)= \Zm(\beta).
\ee
The continuum limit $\as\to0$ is taken at fixed $\xi$.
The  factor $Z_{\sigma,\tau}^\pm$ are such that, for instance,
 $\<\sum_x \Theta_{00}(x)\>{\rightarrow} \<\int d^4x \, \theta_{00}(x)\>$.
The continuum limit of thermodynamic potentials is obtained according to
\ba
\Fp(\betas,\betat,\Nt)  \equiv  \Nt^4\xi^{-3}(\< \Theta\>_T -\<\Theta\>_{0} )   
&\stackrel{\as\rightarrow0}{\rightarrow}  &  \frac{\epsilon-3P}{T^4} \equiv \fp(T)
 \\
\Fm(\betas,\betat,\Nt) \equiv  \frac{4}{3} \Nt^4\xi^{-3} \< \Theta_{00} \> 
&\stackrel{\as\rightarrow0}{\rightarrow} & 
 \frac{\epsilon+P}{T^4} \equiv \fm(T),
\ea
where the leading corrections are ${\rm O}(\as^2)$.

\subsection{Thermodynamics and normalization of $\theta$ and $\theta_{\mu\nu}$}
In this section we relate the normalization factors $Z_{\sigma,\tau}^\pm$ 
to derivatives of the bare parameters with respect to physical parameters.
We start from the thermodynamic relations
\be
\epsilon= -\frac{1}{L^3} \frac{\partial \log \ovl Z}{\partial L_0},\qquad
p= \frac{1}{L_0}\frac{\partial \log \ovl Z}{\partial L^3}
\ee
where 
\be
\log\ovl Z(\betas,\betat,\Ns,\Nt)= \log Z(\betas,\betat,\Ns,\Nt)
  - \frac{\Nt}{\Ntref}\log Z(\betas,\betat,\Ns,\Ntref).
\ee
The subtraction, which sets the  free energy $F=-T\log\ovl Z$ to zero
at a reference temperature $\Tref=1/(\Ntref\at)$, is necessary in quantum field theory. 
On a $\xi=1$ lattice a common choice
is $\Ntref=\Ns$, which implies that $\Tref=0$ in the thermodynamic limit.
We can combine the equations 
\[ \frac{\partial\log \ovl Z}{\partial\log\as} = 0 \qquad {\rm and} \qquad 
   \frac{\partial\log \ovl Z}{\partial\log\xi} = 0  \]
into 
\ba
(\epsilon-3P)\as^3\at &=& 
 \dBsdlogas \<\Ss\>_{T-0} + \dBtdlogas \<\St\>_{T-0}  \\
\threequarter(\epsilon+P)\as^3\at &=& - 
 \Big(\dBsdlogxi +\frac{1}{4} \dBsdlogas\Big) \<\Ss\> - 
\Big(\dBtdlogxi +\frac{1}{4}\dBtdlogas \Big)\<\St\>
\ea
From here we read off the normalization factors of $\Theta$ and $\Theta_{00}$:
\ba
\xi^3\Zps = \frac{\partial\beta_{\sigma}}{\partial\log \as},
&\qquad&
\xi^3\Zpt = \frac{\partial\beta_{\tau}}{\partial\log \as},
\la{eq:Zp}
 \\
\xi^3\Zms = - \dBsdlogxi - \frac{1}{4}\dBsdlogas,
&\qquad &
\xi^3\Zmt = ~ \dBtdlogxi +\frac{1}{4} \dBtdlogas.
\ea
Since, by Euclidean symmetry, $\Zms\stackrel{\xi=1}{=}\Zmt$,
we have the equalities
\ba
\frac{\partial(\betas+\betat)}{\partial\log\xi} 
&\stackrel{\xi=1}{=} &  - \frac{1}{2} \dBdloga,  \\
\frac{\partial(\betat-\betas)(\as,\xi)}{\partial\log\xi}
&\stackrel{\xi=1}{=}& 2\beta Z(\beta)   .
\la{eq:Zb}
\ea
We discuss a different choice of bare parameters often
used in numerical simulations in appendix A.
\subsection{Derivatives of $Z_{\sigma,\tau}^\pm$ at $\xi=1$ \la{sec:dZ}}
At  $\xi=1$, $(\partial_{\betas} + \partial_{\betat})$ becomes $d/d\beta$.
Using
\ba
\left(\begin{array}{l@{~~~}r}
\frac{\partial\log\as}{\partial\betas} &  \frac{\partial\log\as}{\partial\betat} \\
\frac{\partial\log\xi}{\partial\betas} &  \frac{\partial\log\xi}{\partial\betat} 
\end{array}\right)
\stackrel{\xi=1}{=} \frac{1}{2\beta Z(\beta) \dBdloga}
\left(\begin{array}{l@{~~~}r}
\frac{\partial\betat}{\partial\log\xi}  & -\frac{\partial\betas}{\partial\log\xi} \\
-\frac{\partial\betat}{\partial\log\as} &   \frac{\partial\betas}{\partial\log\as}
\end{array}\right),
\ea
one easily obtains the relations
\ba
\frac{1}{2}\Big(\frac{\partial}{\partial\betas}- \frac{\partial}{\partial\betat}\Big) (\Zps + \Zpt) 
&\stackrel{\xi=1}{=}& 3\dBdloga \frac{1}{\beta Z(\beta)},
\nn
 \frac{1}{2}\Big(\frac{\partial}{\partial\betas}- \frac{\partial}{\partial\betat}\Big) (\Zps- \Zpt) 
 &\stackrel{\xi=1}{=}& \dBdloga \frac{\partial_\beta(\beta Z(\beta))}{\beta Z(\beta)} .
\la{eq:Zprel}
\ea
We shall need these relations in the next section. 
Similarly we introduce the quantities
\be
\lambda^{\pm}_{00}(\betas,\betat) \equiv 
\frac{1}{2}\Big(\frac{\partial}{\partial\betas}- \frac{\partial}{\partial\betat}\Big) (\Zms \pm \Zmt).
\la{eq:l00}
\ee
At $\xi=1$ they evaluate to
\ba
\lambda^{+}_{00}(\beta) 
&=& 3-\frac{1}{2}\dBdloga \Big[\frac{1}{\beta}+ \frac{dZ}{Zd\beta} \Big]
+\frac{1}{2\beta Z(\beta)} \frac{\partial^2(\betas-\betat)}{\partial(\log\xi)^2}
,\\
{\beta Z(\beta)\lambda^{-}_{00}(\beta)}{} 
&=& \frac{1}{2}
\Big[-\frac{1}{8}\frac{d^2\beta}{d(\log a)^2}+ 
\frac{\partial^2(\betas+\betat)}{\partial(\log\xi)^2}\Big].
\ea
These derivatives thus depend on second derivatives with respect to $\xi$.
In appendix B we obtain the leading order values of $\lambda^{+}_{00}(\beta)$
in $g_0^2$.
\section{Derivation of the sum rules}
We now derive the sum rules, neglecting ${\rm O}(a^2)$ 
discretization errors, but without using perturbative approximations to 
normalization factors such as $\dBdloga$ and $Z(\beta)$.

\subsection{Derivation  on the isotropic lattice}
We consider a renormalization group invariant (RGI) quantity $f(a,T)$, 
which is obtained as the continuum limit of a function $F(\beta,\Nt)$ 
of the bare parameters.
The renormalization group equation $a\partial f/\partial a=0$ implies
\be
T\frac{\partial f}{\partial T} = -\dBdloga \frac{\partial F}{\partial\beta}
\la{eq:sr_iso}
\ee
We have used 
\[ ad\Nt /da= -\Nt \qquad {\rm and} \qquad \Nt \partial_{\Nt} F = - T\partial_T f. \]
In particular, we can apply this equation to  $\Fpm(\beta,\Nt)$, since
they are RGI quantities (see \eq\ref{eq:Fp}, \ref{eq:Fm}).
For the case of $\Fp$, we obtain
\be
a^{-4}\< {\txts\sum_x} \,\Theta(x) \Theta(0) \>^c_T - a^{-4}\< {\txts\sum_x}\,\Theta(x) \Theta(0) \>^c_0
= T^5 \partial_T \frac{\epsilon-3P}{T^4}
+{\txts \frac{d^2\beta}{d(\log a)^2} 
\frac{1}{d\beta/d\log a}}  ~ (\epsilon-3P).
\la{eq:sr_iso_p}
\ee
This sum rule was first derived in~\cite{kapu1}  in the continuum,
in which case the second term on the right-hand side is absent. 
Indeed, the factor multiplying $(\epsilon-3P)$
behaves asymptotically as $2b_1g_0^4$ at small bare coupling.
Consider next the case of $\Fm$; the sum rule reads
\be
\frac{4}{3a^4}\, 
\< {\txts\sum_x}\, \Theta(x)  \Theta_{00}(0) \>^c_T = T^5\partial_T \frac{\epsilon+P}{T^4}
 + {\txts\dBdloga \frac{\partial_\beta(\beta Z(\beta))}{Z(\beta)\beta}}  ~ (\epsilon+P).
\la{eq:sr_iso_m}
\ee
Note that the left-hand side vanishes by Euclidean symmetry at $T=0$. The factor multiplying
$(\epsilon+P)$ in the second term on the right-hand side vanishes in the continuum limit as $(-2b_0g_0^2)$.

\subsection{Derivation  on the anisotropic lattice}
For any RGI quantity $f(\as,\xi,T)$,  $\as \partial_{\as}f  = 0$ and 
$\xi \partial_{\xi}f=  0$ respectively imply
\be
T\frac{\partial f}{\partial T} 
\left( \begin{array}{c}  -1 \\ 
                          1  \end{array}\right)
= \left(  \begin{array}{c@{\quad}c} \dBsdlogas & \dBtdlogas \\
                                    \dBsdlogxi &  \dBtdlogxi \end{array}\right)
 \left(  \begin{array}{c} \dFdBs \\
                         \dFdBt    \end{array}\right)            
\ee

We have used 
\[ \as\partial_{\as}\Nt = -\Nt \qquad {\rm and} 
\qquad \Nt \partial_{\Nt} F(\betas,\betat,N_\tau) = - T\partial_T f. \]
From now on we  evaluate the expression at the isotropic point $\betas=\betat$.
The determinant of the matrix is then
\be
\Delta = 2\beta Z(\beta)\dBdloga.
\ee
Taking suitable linear combinations, we obtain the two equations
\ba
-T\frac{\partial f}{\partial T} &=& \dBdloga \left(\dFdBs + \dFdBt \right) \la{eq:an_sr1}\\
-\frac{3}{4}T\frac{\partial f}{\partial T} &=& \beta Z(\beta) \left(\dFdBs - \dFdBt \right).
\la{eq:an_sr2}
\ea
The first relation is equivalent to \eq\ref{eq:sr_iso} derived on the isotropic lattice,
since $\frac{d}{dx}f(x,x) = (\partial_y +\partial_z)f|_{y,z=x}$ for  a general
function of two variables $(y,z)$.
We therefore focus on the second relation in the following. 
%

The observables $\fpm$ are RGI quantities. Consider first $\fp(T)$. 
Using \eq\ref{eq:Zprel} and the thermodynamic relations $T\partial_T p =\epsilon+P$ and   
$(\epsilon-3P)/T^4=T\partial_T(p/T^4)$,  
\eq\ref{eq:an_sr2} leads  to 
\eq\ref{eq:sr_iso_m} derived on the isotropic lattice.

We now apply \eq\ref{eq:an_sr2} to $\fm(T)$.
We obtain a new sum rule,
\be
a^4\< {\txts\sum_x}\,\Theta_{00}(x)\Theta_{00}(0)\>^c_T 
-  \beta Z(\beta)\lambda^{-}_{00}(\beta)a^{-4} \<S_+\>_T 
= {\txts\frac{3}{4}} \lambda^{+}_{00}(\beta) (\epsilon+P) 
+ {\txts\left(\frac{3}{4}\right)^2} T^5 \partial_T \frac{\epsilon+P}{T^4}.
\la{eq:an_sr_m}
\ee
The quantities $\lambda^{\pm}_{00}(\beta)$ are defined in \eq\ref{eq:l00}.
Since the right-hand side of \eq\ref{eq:an_sr_m} 
manifestly has a finite continuum limit, this equation implies 
that the short-distance quartic divergence of the integrated correlator is compensated 
by the quartic divergence ($a^{-4}$) of the expectation value of the trace anomaly,
\be
\<\beta S_+\>_T ~ = ~ \frac{3}{2}d_A ~(1+{\rm O}(g_0^2))
~~~~~~~~~~~(d_A\equiv N^2-1).
\ee

\subsection{Contact terms in two-point functions of the Hamiltonian}
The $\<\theta\theta\>$, $\<\theta_{00}\theta_{00}\>$ and $\<\theta_{00}\theta\>$ correlators 
are related at vanishing spatial momentum 
because the Hamiltonian operator $\int d^3{\bf x}T_{00}$ has simple correlation functions:
\be
\<{\txts\int} d^3{\bf x}~ T_{00}(x_0,{\bf x})~ {\cal O} \>_T^c = 
T^2 \partial_T \<{\cal O}\>_T + A_{\cal O}(T)~ \delta(x_0)
\la{eq:T00}
\ee
for any local operator ${\cal O}$.
The delta function arises because the Hamiltonian operator applied on
transfer-matrix eigenstates with energies at the cutoff scale does not yield
the expected matrix elements; for instance off-diagonal matrix elements are expected
to appear in general.

The sum rules (\eq \ref{eq:sr_iso_p}, \ref{eq:sr_iso_m}, \ref{eq:an_sr_m})
determine the contact terms  $A_{\theta_{00}}$ and $A_{\theta}$:
\ba
A_{\theta_{00}} &=& \frac{\lambda^-_{00}(g_0) Z(g_0)}{g_0^2 {\txts dg_0^{-2}/d\log a}} \<\theta\>_T
   + 3 \left(\quarter\lambda^+_{00}(g_0)+
{\frac{g_0^2}{16}} \frac{dg_0^{-2}}{d\log a}\left[1-\frac{g_0^2}{Z}\frac{dZ}{dg_0^2}\right] -1 \right) (\epsilon+P), \\
A_{\theta} &=& \left(\frac{1}{4} \frac{d^2g_0^{-2}}{d(\log a)^2} 
               \frac{1}{\txts dg_0^{-2}/d\log a}-1\right)(\epsilon-3P) 
+\frac{3}{4}  g_0^2 \frac{dg_0^{-2}}{d\log a} \left[1-\frac{g_0^2}{Z}\frac{dZ}{dg_0^2}\right] (\epsilon+P) \nn
&& +\quarter \<{\txts\int} d^4 x ~\theta(x)\theta(0)\>_0^c.
\ea
The contact term of $\<T_{00}T_{00}\>$ is then given by 
$A_{T_{00}}=A_{\theta_{00}}+\quarter A_{\theta}$. Note that the contact terms 
have a quartically divergent contribution, plus finite, temperature-dependent contributions.
\section{Sum rules in the continuum}
Taking the bare coupling $g_0^2\ll 1$ in \eq 
(\ref{eq:sr_iso_p}, \ref{eq:sr_iso_m}, \ref{eq:an_sr_m}) yields the 
following continuum sum rules:
\ba
\< {\txts\int} d^4x \,\theta(x) \theta(0) \>^c_T - \< {\txts\int} d^4x \,\theta(x) \theta(0) \>^c_0
&=& T^5 \partial_T \frac{\epsilon-3P}{T^4}\la{eq:src1} \\
 \< {\txts\int} d^4x\, \theta(x)  \theta_{00}(0) \>^c_T &=& 
{\txts\frac{3}{4}} T^5\partial_T \frac{\epsilon+P}{T^4}
\la{eq:src2}
  \\
\< {\txts \int}d^4x\,\theta_{00}(x)\theta_{00}(0)\>^c_T 
+  {\txts\frac{\lambda^-_{00}}{2b_0g_0^2}} \<\theta\>_T 
&=& {\txts\frac{3}{4}} \lambda^{+}_{00} (\epsilon+P) 
+ {\txts\left(\frac{3}{4}\right)^2} T^5 \partial_T \frac{\epsilon+P}{T^4}.
\la{eq:src3}
\ea
The coefficients $\lambda_{00}^\pm$ are now to be taken at $g_0=0$, where they are pure, finite numbers.
We compute these numbers in appendix B, see \eq (\ref{eq:l00-tl}) and (\ref{eq:l00+tl}).
The  calculation of $\lambda_{00}^+$ suggests that the latter is independent of 
the regularization used. If true, this would mean that the regularization dependence
cancels entirely between the two terms on the left-hand side of this equation.
It would be useful to derive \eq\ref{eq:src3} in a different regularization to confirm this.

The difference of relation (\ref{eq:src3}) between finite and zero-temperature gives
\[
 \< {\txts \int}d^4x\,\theta_{00}(x)\theta_{00}(0)\>^c_T 
 - \< {\txts \int}d^4x\,\theta_{00}(x)\theta_{00}(0)\>^c_0
+  {\txts\frac{\lambda^{-}_{00}}{2b_0g_0^2}} (\epsilon-3P) =
{\txts\frac{3}{4}} \lambda^{+}_{00} (\epsilon+P) 
+ {\txts\left(\frac{3}{4}\right)^2} T^5 \partial_T \frac{\epsilon+P}{T^4}.
\]
This relation shows that even after subtraction of the quartic divergence, a temperature-dependent 
logarithmic divergence remains in the susceptibility of $\theta_{00}$.

\section{Sum rules and cutoff effects on $\epsilon$ and $P$}
%
The idea to remove the leading cutoff effects on physical quantities 
by using lattice sum rules was proposed in~\cite{hmsr1}. Here we show that it can be 
applied to thermodynamic potentials. Consider for instance $(\epsilon-3P)/T^4$.
On a $\xi=1$ lattice, this quantity is obtained by taking the $\Nt\rightarrow\infty$ limit of 
\be
\varphi(\Nt) \equiv \Fp(\beta(\Nt),\Nt),
\ee
where $\beta(\Nt)$ is tuned so that $(\Nt a)$ is constant and $\Fp$ was 
defined in \eq\ref{eq:Fp}. Following the steps of section 3, we can evaluate
\ba
\frac{d\varphi}{d\log \Nt}&=&\frac{\partial \Fp}
{\partial\log\Nt}- \dBdloga \frac{\partial \Fp}{\partial\beta} \nn
&=& \frac{\partial \Fp}{\partial\log\Nt}
- \frac{d^2\beta}{d(\log a)^2} 
\varphi(\Nt)+\Nt^4 \< {\txts\sum_x}\Theta(x)\,\Theta(0)\>^c_{T-0}
\ea
Thus the cutoff effects can be evaluated in Monte-Carlo simulations at fixed $\beta$.
The first term is itself unambiguous only up to ${\rm O}(a^2)$ 
if a \emph{symmetric} difference scheme is used, and ${\rm O}(a)$ if not~\cite{hmsr1}.
It requires performing a simulation at a second value of $\Nt$.
Thus in total three simulations are required (for instance with the number of points 
in the time direction set to $\Nt$, $\Nt+1$, and $\Ns$ for the zero-temperature subtractions).
Choosing a different couple $(\beta(\Nt'),\Nt')$ tuned to the same temperature
requires four simulations in total and provides essentially the same information 
(unless $\Nt'$ is much larger than $\Nt$, but in practice, typical values are
$\Nt=6$ and $\Nt '=8$). If one follows both strategies, 
one can check how close $\varphi(\Nt')$ is from
\[
\varphi(\Nt) + \frac{1}{2} \frac{d\varphi}{d\log\Nt} \left(1-(\Nt/\Nt')^2\right).
\]
If $\Nt$ is large enough, $\varphi$ is in the regime where 
O($a^2$) effects dominate over higher order cutoffe effects
and $\varphi(\Nt')$ will be numerically consistent with this expression.
In general, this provides a way of testing whether $\varphi$ is in this regime
without having to perform simulations at $\Nt''>\Nt'$.
Since the cost of finite-temperature calculations grows with a high power 
of $\Nt$, this information is very precious.


%
\section{Conclusion}
We have derived finite temperature sum rules, valid at finite lattice spacing
up to ${\rm O}(a^2)$ corrections. The main results are \eq
(\ref{eq:sr_iso_p}, \ref{eq:sr_iso_m}, \ref{eq:an_sr_m}), and, 
for the reader interested in continuum results,
\eq (\ref{eq:src1}--\ref{eq:src3}).

As an application of these considerations, we have proposed a way
to check whether thermodynamics calculations are performed 
in the regime where the O($a^2$) cutoff effects dominate over 
higher order cutoff effects, using only two values of $\Nt$.

Further sum rules can be obtained for other RGI quantities. 
Equations \ref{eq:sr_iso} and (\ref{eq:an_sr1},\ref{eq:an_sr2}) can for instance 
be applied to renormalized Polyakov or Wilson loops in order
to study thermal contributions to quark masses, 
and the static potential relevant to 
J/$\psi$ suppression~\cite{laine}. Finally the sum rules can
be generalized to full QCD with commonly used quark actions.

\paragraph{}
This work was supported in part by
funds provided by the U.S. Department of Energy under cooperative research agreement
DE-FG02-94ER40818.

\appendix
\section{A different choice of bare parameters}
Although the set of bare parameters $(\betas,\betat)$ is most convenient to derive sum rules, 
in numerical practice, it is more convenient to parametrize these parameters as 
\be
\betas=\frac{\beta}{\xi_0} \qquad\qquad  \betat=\beta\xi_0.
\ee
In order to take the continuum limit at fixed anisotropy $\xi$, 
the first task of the lattice practitioner is to establish the lines
of constant $\xi$ in the $(\beta,\xi_0)$ plane, so that $\xi_0$ can 
thereafter be viewed as a function of $(\beta,\xi)$.
Secondly the relation between $\beta$ and $\as$ must be worked out 
at fixed anisotropy $\xi$. After this preparatory work, the  set of variables used in practice
is $(\beta,\xi)$. 

The expression (\ref{eq:Zp}) can thus be written as 
\ba
\frac{\partial\beta_{\sigma}(\as,\xi)}{\partial\log \as} 
&=& \frac{1}{\xi_0} \frac{\partial\beta(\as,\xi)}{\partial\log \as} 
\Big[ 1- \frac{\beta}{\xi_0} \frac{\partial\xi_0(\beta,\xi)}{\partial\beta} \Big] \\
\frac{\partial\beta_{\tau}(\as,\xi)}{\partial\log \as}   
  &=& \xi_0\,\frac{\partial\beta(\as,\xi)}{\partial\log \as} 
\Big[ 1+ \frac{\beta}{\xi_0} \frac{\partial\xi_0(\beta,\xi)}{\partial\beta} \Big].
\ea
Similarly, using 
 \[
 \frac{\partial\beta(\as,\xi)}{\partial\log\xi}= 
  -\frac{\partial\beta(\as,\xi)}{\partial\log\as} \frac{\partial\log\as(\beta,\xi)}{\partial\log\xi}
 =-\frac{\partial\beta(\as,\xi)}{\partial\log\as}
 \frac{\partial\log\xi_0(\beta,\xi)}{\partial\log\xi} \frac{\partial\log\as(\beta,\xi_0)}{\partial\log\xi_0},
 \]
we obtain
\ba
\frac{\partial\betas(\as,\xi)}{\partial\log\xi}&=& -\frac{\beta}{\xi_0}~
\frac{\partial\log\xi_0(\beta,\xi)}{\partial\log\xi} ~~\times  \\
&&\Big[ 1+\frac{\partial\beta(\as,\xi)}{\partial\log\as}\frac{\partial\log\as(\beta,\xi_0)}{\partial\log\xi_0}
\Big(\frac{1}{\beta}- \frac{\partial\log\xi_0(\beta,\xi)}{\partial\beta}  \Big) \Big]. \nn
\frac{\partial\betat(\as,\xi)}{\partial\log\xi}&=&
\xi_0\beta ~\frac{\partial\log\xi_0(\beta,\xi)}{\partial\log\xi} ~~\times  \\
&&\Big[ 1- \frac{\partial\beta(\as,\xi)}{\partial\log\as}\frac{\partial\log\as(\beta,\xi_0)}{\partial\log\xi_0}
\Big(\frac{1}{\beta}+\frac{\partial\log\xi_0(\beta,\xi)}{\partial\beta}\Big)\Big]. \nonumber
\ea
These expressions suggest how to determine $\frac{\partial\beta_{\sigma,\tau}(\as,\xi)}{\partial\log\xi}$
non-perturbatively. Since, by Euclidean symmetry, $\Zms\stackrel{\xi=1}{=}\Zmt$ and 
$\frac{\partial\xi_0(\beta,\xi)}{\partial\beta}\stackrel{\xi=1}{=} 0$,
we have the equalities
\ba
\frac{\partial\beta(\as,\xi)}{\partial\log\xi}
&\stackrel{\xi=1}{=}&  - \frac{1}{4} \dBdloga, \\
Z(\beta)
& \stackrel{\xi=1}{=} &
\frac{\partial\xi_0(\beta,\xi)}{\partial\xi}.
\la{eq:Zb2}
\ea

 \section{Leading-order computation of $\lambda_{00}^{\pm}$}
In this appendix we calculate the coefficients $\lambda_{00}^{\pm}(\beta)$
defined in \eq\ref{eq:l00}. 
Using the standard notation $\hat p_\mu= 2\sin (p_\mu/2)$,
we define the dimensionless integrals
\ba
 I_\sigma(\xi_0^2,\Nt)&=& \frac{1}{\Nt}\sum_{p_0}\int_{-\pi}^{\pi} \frac{d^3p}{(2\pi)^3} ~
 \frac{\hat{p}_1^2 }{ \xi_0^2\hat p_0^2+{\txts \sum_k}\hat{p}_k^2 } \\
I_\tau(\xi_0^2,\Nt)&=& \frac{1}{\Nt}\sum_{p_0}\int_{-\pi}^{\pi} \frac{d^3p}{(2\pi)^3} ~
 \frac{\xi_0^2\hat{p}_0^2 }{ \xi_0^2\hat p_0^2+{\txts \sum_k}\hat{p}_k^2 } 
\ea
The variable $p_0$ takes the values $2\pi k/\Nt$ for $0\leq k<\Nt$.
One finds that $I_{\sigma}(1,\infty)=I_{\tau}(1,\infty)=1/4$ and 
\be
3\frac{\partial I_{\sigma}}{\partial\xi_0^2}(\xi_0^2=1,\infty)
=  -\frac{1}{4} + \int_{-\pi}^{\pi} \frac{d^4p}{(2\pi)^4} 
 \frac{\hat p_0^4}{(\hat p_0^2+{\txts \sum_k}\hat{p}_k^2 )^2}  = -0.154933\dots
\ee


\subsection{$\lambda_{00}^{-}$}
Notice first that $\lambda_{00}^{-}(\beta)$ can be rewritten
\be
\lambda_{00}^{-}(\beta)=
\frac{1}{2}  Z^-(\beta)
\Big(\frac{\partial}{\partial\betas}- \frac{\partial}{\partial\betat}\Big) 
\log\left(\frac{\Zms}{\Zmt}\right).
\ee
The matrix elements of $\theta_{00}$ on physical states are RGI quantities.
On a $\xi=1$ lattice, 
\be
\<\Omega|\theta_{00}|\Omega\>=0, 
\la{eq:00vac}
\ee
as a consequence of the Euclidean symmetry on an $\Ns=\Nt=\infty$ lattice.
Therefore \eq\ref{eq:00vac} must be satisfied  also on a $\xi\neq1$
lattice. This condition determines the ratio ${\Zms}/{\Zmt}$:
\be
\frac{\Zms}{\Zmt}= \frac{\<S_\tau\>_0}{\<S_\sigma\>_0}.
\la{eq:cond1}
\ee
At leading order on an $\Nt\times\infty^3$ lattice,
\ba
\frac{\betat}{d_A} \< S_\tau \>_T &=& \frac{3}{2} \left[I_\tau(\xi_0^2,\Nt)+I_\sigma(\xi_0^2,\Nt)\right]
  = \frac{3}{2}-3I_\sigma(\xi_0^2,\Nt), \\
\frac{\betas}{d_A} \< S_\sigma \>_T &=& 3 I_\sigma(\xi_0^2,\Nt)
~~~~~~~~~~~~~~~~~~~~~~~(T=(a_\tau\Nt)^{-1}),
\ea
which leads to
\be
 \lambda^-_{00} =  1+8~\frac{\partial I_\sigma}{\partial\xi_0^2}(\xi_0^2=1,\infty)=0.586844\dots
\la{eq:l00-tl}
\ee

\subsection{$\lambda_{00}^{+}$}
A second physics condition (in addition to \eq\ref{eq:cond1}) 
is necessary in order to fix $Z^-_\sigma$ and $Z^-_\tau$ 
separately and therefore to determine $\lambda_{00}^{+}$. 
We impose the condition 
\be
\frac{\Nt^4}{\xi^3}\<\Theta_{00}\>_T = \frac{3}{4}(\epsilon+P)/T^4
\la{eq:cond2}
\ee
This leads to the expressions
\ba
Z^-_\sigma &=& \left[\frac{3}{4}\frac{\epsilon+P}{d_AT^4}\right]~\frac{\betas}{\Nt^4W_T(\xi_0^2)}~
\<\betat S_\tau/d_A\>_0 \nn
Z^-_\tau  & = &\left[\frac{3}{4}\frac{\epsilon+P}{d_AT^4}\right]~\frac{\betat}{\Nt^4W_T(\xi_0^2)}~
\<\betas S_\sigma/d_A\>_0
\la{eq:zst}
\ea
where 
\be
W_T(\xi_0^2)\equiv \<\betat S_\tau/d_A\>_0 ~\<\betas S_\sigma/d_A\>_T ~-~ 
\<\betat S_\tau/d_A\>_T ~\<\betas S_\sigma/d_A\>_0.
\ee
Cutoff effect due to finite $\Nt$ can be removed by taking the limit $\Nt\to\infty$.
Expressions \eq\ref{eq:zst} in principle allow for a non-perturbative determination of $Z^-_{\sigma,\tau}$, 
but at tree level we shall use the Stefan-Boltzmann expression $\pi^2d_A/15$ for the right-hand side
of \eq\ref{eq:cond2}.  In that approximation we have
\be
W_T(\xi_0^2) \stackrel{LO}{=} \frac{9}{2} \left[ I_\sigma(\xi_0^2,\Nt) -  I_\sigma(\xi_0^2,\infty)\right].
\ee
For $\xi_0^2=1$, we know that both $Z_\sigma^-$ and $Z_\tau^-$ are equal to $\beta$
at leading order. Thus 
\be
\lim_{\Nt\to\infty}~ \Nt^4 ~W_T(1) = \frac{\pi^2}{20}.
\ee
Because $\lim_{\Nt\to\infty}\Nt^4 W_T(\xi_0^2)$ is a continuum limit, $\hat p$ can be replaced by $p$ 
and one then finds that 
\be
\lim_{\Nt\to\infty}~ \Nt^4 ~W_T(\xi_0^2) =\xi_0^3 \lim_{\Nt\to\infty}~ \Nt^4 ~W_T(1) =\frac{\pi^2\xi_0^3}{20}.
\ee
One then straightfowardly obtains
\be
\lambda^+_{00} = 6.
\la{eq:l00+tl}
\ee

{\small

}
\end{document}